\documentclass[12pt,preprint]{aastex}

\slugcomment{To appear in ApJ}

\newcommand{\gI}{$g\!-\!I$}
\shorttitle{The Formation of Spiral Spheroids}
\shortauthors{Maybhate et al.}

\begin{document}


\title{The Formation of Spheroids in Early-Type Spirals: Clues From
Their Globular Clusters\altaffilmark{1}} 
\author{Aparna Maybhate\altaffilmark{2}, Paul Goudfrooij\altaffilmark{2},
Rupali Chandar\altaffilmark{3}, and Thomas H. Puzia\altaffilmark{4}}
\altaffiltext{1}{Based on observations with the NASA/ESA {\it Hubble Space
    Telescope}, obtained at the Space Telescope Science Institute, which is
  operated by the Association of Universities for Research in Astronomy, Inc.,
  under NASA contract NAS5-26555} 
\altaffiltext{2}{Space Telescope Science Institute, 3700 San Martin Drive, Baltimore, MD 21218}
\altaffiltext{3}{Department of Physics and Astronomy, The University of Toledo, Toledo, OH 43606}
\altaffiltext{4}{Herzberg Institute of Astrophysics, 5071 West Saanich Road, Victoria, BC V9E 2E7, Canada}

\begin{abstract}
We use deep {\em Hubble Space Telescope} images taken with the
Advanced Camera for Surveys (ACS) in the F475W and
F814W filters to investigate the globular cluster systems in four
edge-on Sa spiral galaxies covering a factor of 4 in luminosity.
The specific frequencies
of the blue globular clusters in the galaxies in our sample fall in
the range 0.34 -- 0.84, similar to typical values found for later-type
spirals. The number of red globular clusters associated with the
bulges generally increases with the bulge luminosity, similar to what
is observed for elliptical galaxies, although the specific frequency
of bulge clusters is a factor
of 2-3 lower for the lowest luminosity bulges than for the higher
luminosity bulges. We present a new empirical relation between the
fraction of red globular clusters and total bulge luminosity based on
the elliptical galaxies studied by ACSVCS (ACS Virgo Cluster Survey), and discuss how this
diagram can be used to assess the importance that dissipative
processes played in building spiral bulges.  Our results suggest a
picture where dissipative processes, which are expected during gas-rich major mergers,  were more important for
building luminous bulges of Sa galaxies, whereas secular
evolution may have played a larger role in building lower-luminosity
bulges in spirals.

\end{abstract}

\section{Introduction}
\label{s:intro}

The assembly history of spiral galaxies remains one of the most pressing
questions in astrophysics today \citep[e.g.,][]{kormendy04}. In particular, we
do not have a clear picture of the formation mechanism for the central bulges
of spiral galaxies.
In the early universe, galactic evolution was primarily driven by
violent merging and dissipative collapse. Mergers are accompanied by
dissipation and star formation and are believed to make classical bulges and elliptical
galaxies.
Are bulges of spirals simply ``small ellipticals'', having
formed through vigorous star formation through short time scale
dissipative processes or merging during the early universe , with
disks assembled later on?
Or is bulge formation through secular evolution of inner disk stars a
more common mechanism? Secular evolution is a slow process and involves the
redistribution of disk material. In this scenario, material is
driven into the galaxy center by means of a bar instability
\citep[e.g.,][]{pfenniger90,sheth05} or a minor interaction with a small
satellite galaxy (up to 10$\%$ of the main galaxy mass; see \citet{pfenniger99}),
triggering low-level star formation.
As a result, galaxies would evolve from late-types to early-types with the growth of
their bulges. A relatively new and powerful way to probe these fundamental questions is the 
study of the spheroids and their associated globular cluster (GC) systems in
early-type spiral galaxies.

In ``normal'' elliptical galaxies, previous {\it HST\/} studies have
shown that the broad-band color distribution of the GC systems
are more consistent with two subpopulations of star clusters.
Under the assumption that the globular clusters in these systems have ages
greater than a few Gyrs, this color difference indicates the presence of
two GC populations of different metallicities: Blue (metal-poor) and red (metal-rich) clusters 
\citep[e.g.,][]{gebhardt99,kundu01,larsen01,peng06}. The specific
frequency of GCs ($S_N$, the total number of GCs per unit galaxy luminosity) in
``normal'' ellipticals is high, typically in the range 1.5\,--\,4
\citep[e.g.,][]{peng08}.  
The spatial distribution and colors of massive metal-rich (red) GCs closely follow
that of the spheroidal light (i.e., the ``bulge''), whereas the
metal-poor (blue) GC system is usually more extended 
\citep[``halo-like'',][]{kundu98,dirsch03,puzia04,rhode04,goudfrooij07}. This
is generally interpreted as strong evidence that the formation of elliptical 
galaxies involved the formation of large numbers of such massive metal-rich
GCs, and hence that star formation occurred in a vigorous
and rapid manner (``dissipative  collapse'').  

If the bulges of spirals formed in a manner similar to that in ellipticals,
we expect galaxies with larger bulge-to-total luminosity ratios to be
associated with a larger number of clusters.
At face value, these results seem consistent with
a relatively simple picture in which GC systems of spirals are made up of a
``universal'' halo population of metal-poor GCs that is present in each
galaxy, plus a metal-rich bulge population which grows with the
bulge mass or luminosity. On the other hand, 
in the ``secular evolution'' scenario, galaxies would evolve from late-types to
early-types with the growth of their bulges, but the properties of their GC
systems would remain largely unchanged because low star formation rates do
not allow for the formation of star clusters that are massive enough to
survive a Hubble time \citep[e.g.,][]{bastian08}.

Studies of GC systems in spirals with Hubble types of Sb and later
\citep{goudfroo03} and nearby 
face-on spirals \citep{chandar04} showed that they have
$S_N$ values of  $0.55\pm0.25$, 
i.e. much lower than in ellipticals, whereas 
earlier-type spirals typically have higher $S_N$ values \citep[see
also][]{georgiev10}.
This scenario would indicate that spiral bulges 
are like small ellipticals having formed early on 
through rapid dissipative collapse and now residing at the
center of a large disk which was mostly assembled later.
Based on properties of GC systems in the Milky Way, M31, and M104,
\citet{forbes2001} suggested that red, metal-rich GCs in spirals are
associated with bulges, and are the direct analogs of the red GCs seen in
ellipticals. If this association can be verified and established, this would
provide an important causal link between the formation of metal-rich GCs and
that of spheroidal systems (i.e., bulges and ellipticals).
Thus, a thorough study of GC systems
in spirals can provide important new clues regarding the formation of
bulges/spheroids in spiral galaxies.

In the simple picture described above, the unambiguous detection of a
relationship between bulge luminosity/mass and the specific frequency
of GCs is highly dependent on the results for bulge-dominated spirals
(i.e., Sa galaxies) but studies of GC systems of Sa
galaxies are still rather sparse.  In this paper, 
we test the two basic scenarios for the formation of bulges in Sa galaxies: (1) vigorous star (and cluster)
formation through rapid dissipative processes which could be triggered
by merging and (2) the slow secular evolution accompanied by low levels of
star formation. We present the results of a 
new study of the GC systems of four edge-on, bulge-dominated
(early-type) spiral galaxies.  
We focus on the GC color distributions and the relations between the GC 
specific frequency $S_N$ and host galaxy luminosity as well as the
bulge-to-total luminosity ratio.

\section{Observations and Analysis of HST/ACS Data}
\label{s:obs}

As part of {\it HST} program GO-10594 (PI: Goudfrooij) we selected four nearby,
edge-on, bulge-dominated spiral galaxies to study their GC systems. These
galaxies all have the same morphological type (Sa), and they span a factor of $\sim 4$ 
in luminosity. The basic properties of our target galaxies, including their distance
and absolute magnitude in the $B$ and $K$ bands are given in Table~1.

\begin{figure}
\plotone{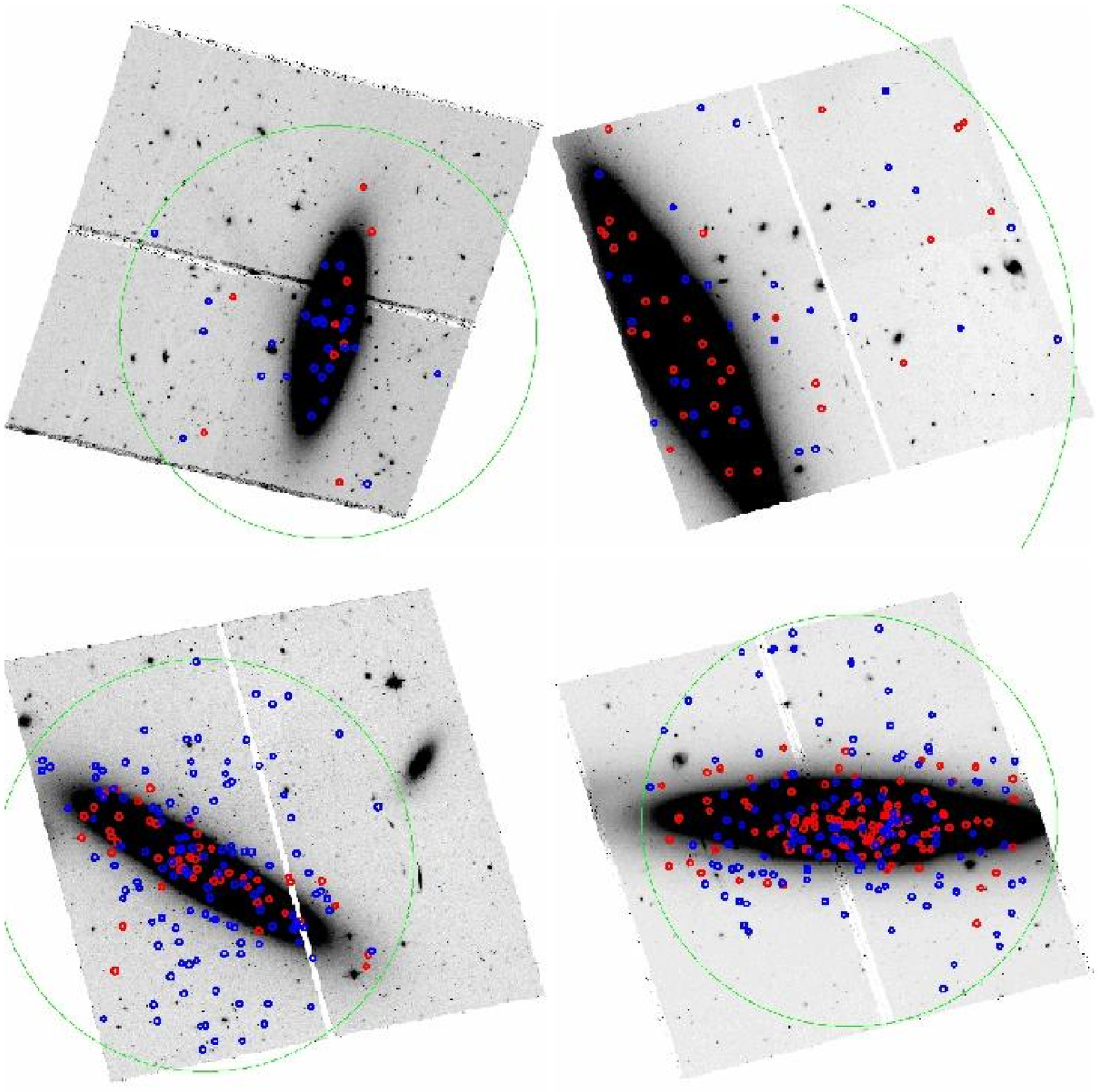}
\caption{Positions of clusters superposed on the $I$-band ACS/WFC images of (clockwise from top left) NGC~5475,
NGC~4710, NGC~4866, and NGC~5308. See text for definitions of red and blue clusters. The green curve is a circle of radius 14 kpc drawn around the galaxy center. North is to the top and East is to the left.}
\label{pos}
\end{figure}

Observations were made using the Wide Field Channel (WFC) of the Advanced Camera
for Surveys (ACS) aboard the {\it Hubble Space Telescope} in the F475W and
F814W filters. The WFC has a 202\arcsec\ $\times$ 202\arcsec\ field of view with
a plate scale of 0\arcsec .05 per pixel. We chose these filters as the best compromise between
high sensitivity and long color baseline.
Total exposure times ranged from 1164 sec to 3820 sec in F475W and from 1000 sec to
2320 sec in F814W. The flatfielded images produced by the {\it HST} pipeline
were run through the task MULTIDRIZZLE, with the background subtraction turned off,
resulting in geometrically corrected images with most cosmic rays removed.
These images are shown in Fig.~\ref{pos}.

Object detection was done following the general procedure described in
\citet{maybhate07}. First, we removed the strong and varying galaxy
light by applying a circular median
filter of radius 15 pixels and dividing the image in each filter by the
square root of this
median image thus providing uniform shot noise characteristics
over the whole frame. Object finding was done on the square root divided images in each
filter. We detected sources in our images using the DAOFIND task in IRAF using a
detection threshold of 4$\sigma$ above the background.

Next, we selected candidate GCs from the source lists by separating them from
foreground stars and background galaxies based on their colors and morphology.
Total magnitudes of these sources were measured and transformed from STMAG to
SDSS {\it g}- and Cousins {\it I} band using the SYNPHOT package within STSDAS
\citep[see][]{maybhate07}. Our final selection of cluster candidates was done 
based on the following criteria:
\begin{itemize}
\item{}1.1 $\leq$ \gI\ $\leq$ 2.2
\item{}$M_I$ brighter than 1.0 mag beyond the expected turn-over of the
luminosity function ($M_{I_{TO}}=-8.46$; \citealt{kundu01})
\item{}1.5 $\leq$ FWHM $\leq$ 5.0 pixels
\item{}0.9 $\leq$ magnitudes between aperture radii of 1 pixel and 3 pixels $\leq$ 1.6
\end{itemize}
We apply criteria for both the FWHM and the concentration index since we found previously
that using only one of these does not fully reject extended sources
(see Fig.~1 in \citet{goudfrooij07}).
We perform artificial cluster experiments to determine the completeness of our samples, by
randomly distributing synthetic point sources with magnitudes between 22.0 and 28.0
over five distinct background levels in the $I$-band image for each
galaxy. Batches of 100 sources each were added and recovered from
each image in brightness intervals of 0.25 mag. See \citet{maybhate07} for
details of this procedure. 

Sources with completeness fractions $\geq 50 \%$ were
selected to make the final list of GCs. The final lists contain 44, 63, 191,
and 273 clusters for NGC 5475, NGC 4710, NGC 5308, and NGC 4866 respectively.

\section{Structural Parameters for Target Galaxies}
\label{s:B/T}

We use the {\it HST} images to determine the
ratio of bulge to total luminosity ($B/T$) for each galaxy.
Bulge parameters were derived for each galaxy using the $I$-band image and the
bulge/disk decomposition method described in \citet{goudfroo03} with one
exception: Rather than fitting an exponential profile (representing a disk)
plus a \citet{deV53} $r^{1/4}$ profile (representing a bulge) to the observed
minor-axis profile of the galaxies, we fit an exponential profile 
plus a \citet{sersic63} profile 
\begin{equation}
\Sigma_{\rm S}(z) = \Sigma_{\rm s,e} \; \exp\left\{-a\;[(z/z_{\rm s,e})^{1/n} -
    1]\right\}
\end{equation}
where $z_{\rm s,e}$ is the S\'ersic effective radius along the minor axis, $a$
is a scale factor, $\Sigma_{\rm s,e}$ is the intensity at 
$z = z_{\rm  s,e}$, and $n$ is the S\'ersic parameter indicating the slope of
the bulge profile. The fit was done by means of an iterative non-linear
minimization routine after making initial
estimates of the fit parameters using bootstrap tests. Areas where dust
extinction is apparent in the image were flagged and ignored during the
fitting process. Minor-axis $z_{\rm s,e}$ values were converted to bulge
effective radii $r_{\rm s,e} = z_{\rm s,e}/\sqrt{1-\epsilon}$ where $\epsilon$
is the ellipticity of the bulge as derived from ellipse fits to the $I$-band
image (using the STSDAS task {\sc ellipse}). 
Total bulge luminosities $B$ were then derived using 
\[ B = 22.7 \;\exp(2n - 0.324 - a) \; \Sigma_{\rm s,e} \; r_{\rm
  s,e}^2 \] 
\citep[cf.][]{ciotti91}.

Since these galaxies do not entirely fit on the ACS/WFC images, it is not
possible to estimate the total $I$-band luminosity using our images alone.
Assuming that the 
photometric growth curves of the target galaxies are the same for $I$ as for
$J$, since these two bands are reasonably close in wavelength, 
we apply the aperture correction from  14\arcsec\
to infinity that is available from the 2MASS catalog to derive the total $I$-band magnitude
for each galaxy. $B/T$ values were determined for each galaxy 
from these values for the total luminosity and the bulge luminosity computed
above.
Resulting values for $r_{\rm s,e}$, $n$, and $B/T$ are listed in 
Table~\ref{t:results}. 

\section{Color Distributions of the Globular Clusters}
\label{s:colordist}

Many GC systems have bimodal color distributions,
which provide important clues to the formation of the host galaxies.
Since most previous studies of GC systems used $V-I$ colors, we transformed
$V-I$ to $g-I$ using the GALEV simple stellar population models \citep{galev}
for an assumed age of 13 Gyr. The GALEV models are calibrated to the ACS
filters and treat the chemical evolution of
the gas and the spectral evolution of the stellar content simultaneously.
They also include absorption as well as emission, both line and continuum.
Using the
results of \citet{larsen01}, we find that typical peaks
for GC systems in E/S0 galaxies are expected to occur at $g-I$ values of
1.35\,--\,1.40 for the blue (metal-poor) clusters \citep[see
also][]{maybhate07,maybhate09} and 1.72\,--\,1.85 for the red (metal-rich) clusters. 
Using the bulge luminosities of our sample galaxies, the linear 
relations between peak $V-I$ color (for both blue GCs and red GCs) and
elliptical galaxy luminosity of \citet{larsen01} are used to predict $g-I$
colors to divide  ``blue'' and ``red'' clusters. These colors are listed in
Table~\ref{t:results}.  
Figure~\ref{pos} shows the positions of the red and blue cluster candidates on
the $I$-band ACS image.  

\begin{figure*}
\plottwo{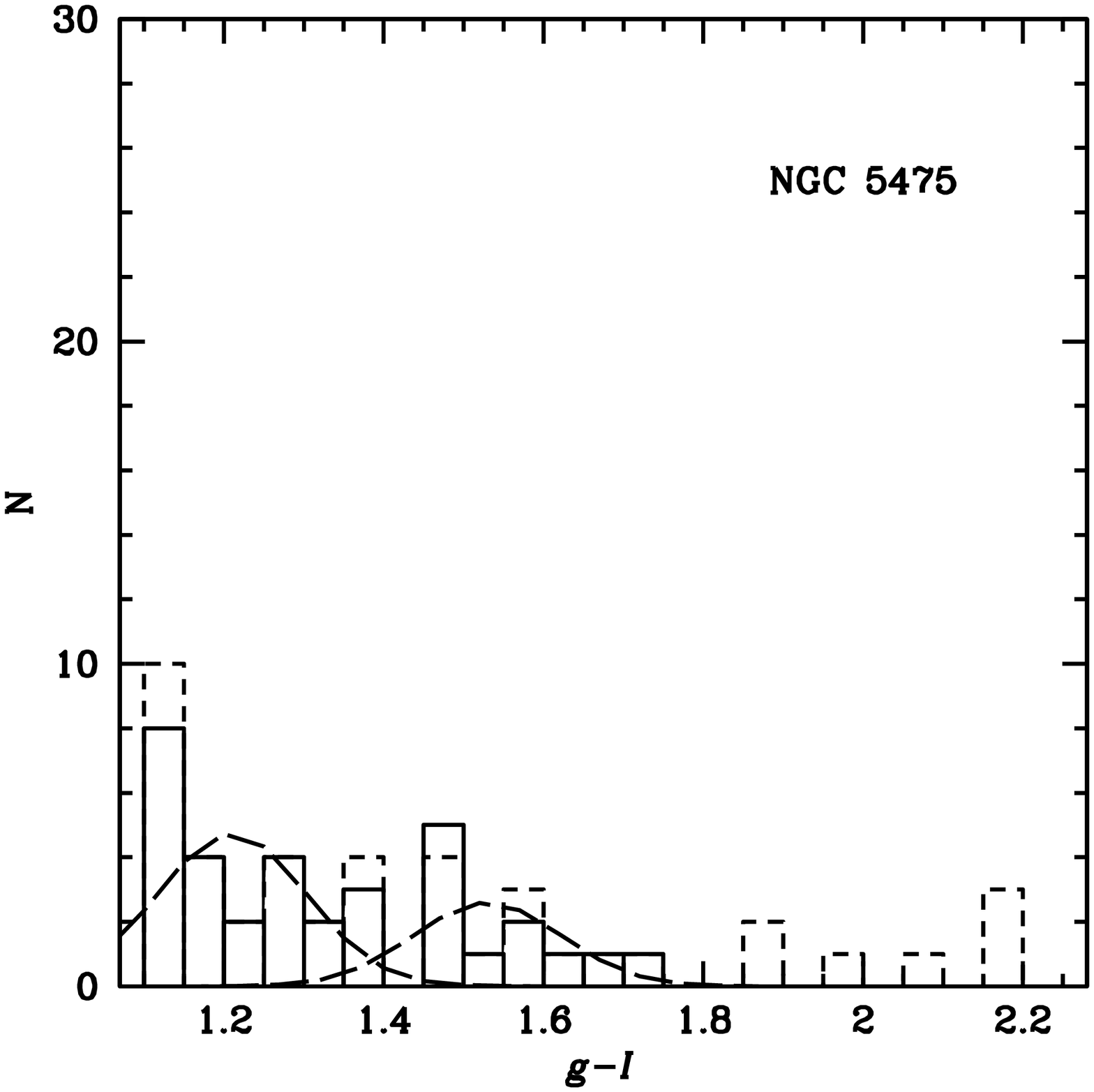}{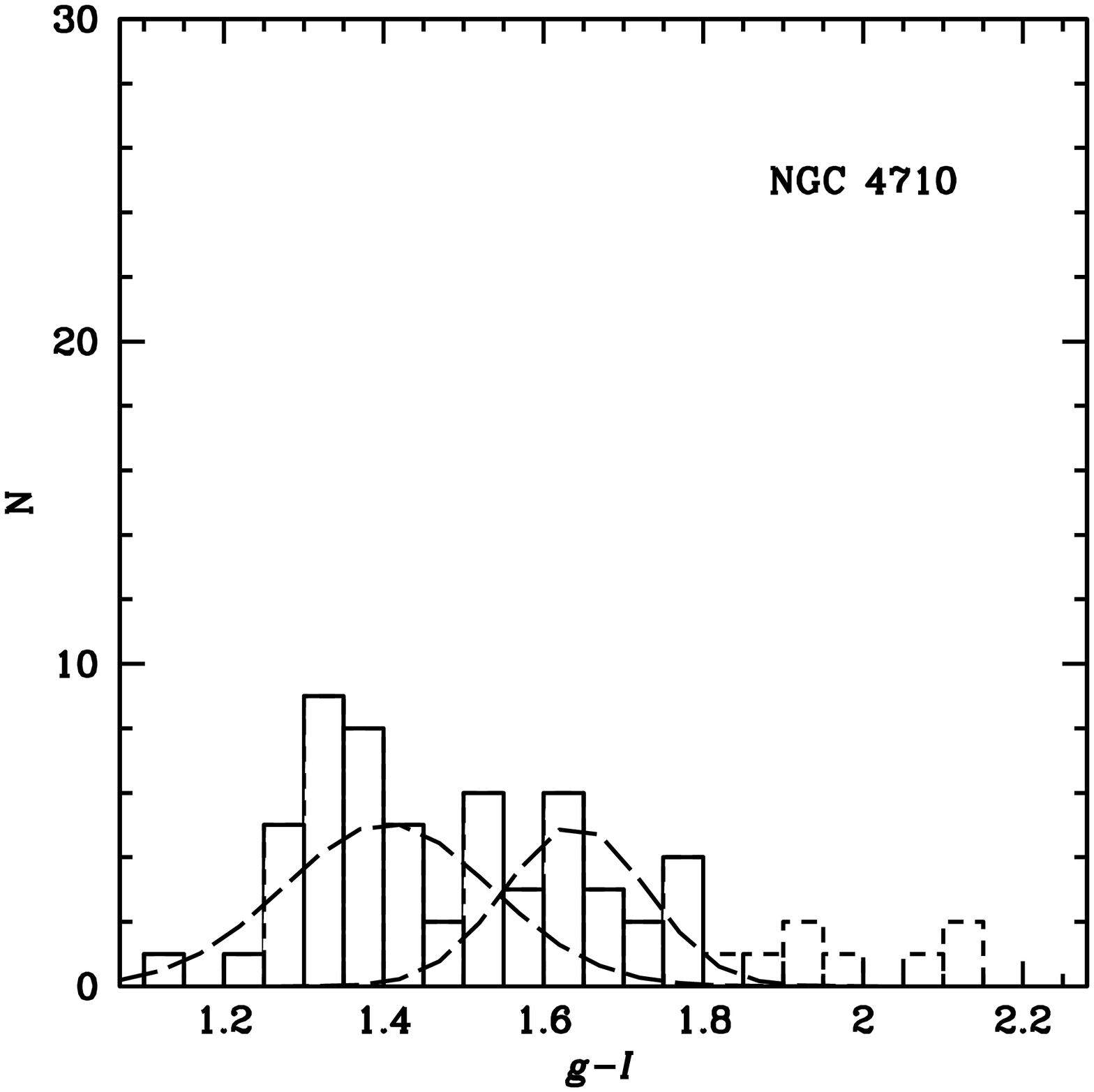}
\plottwo{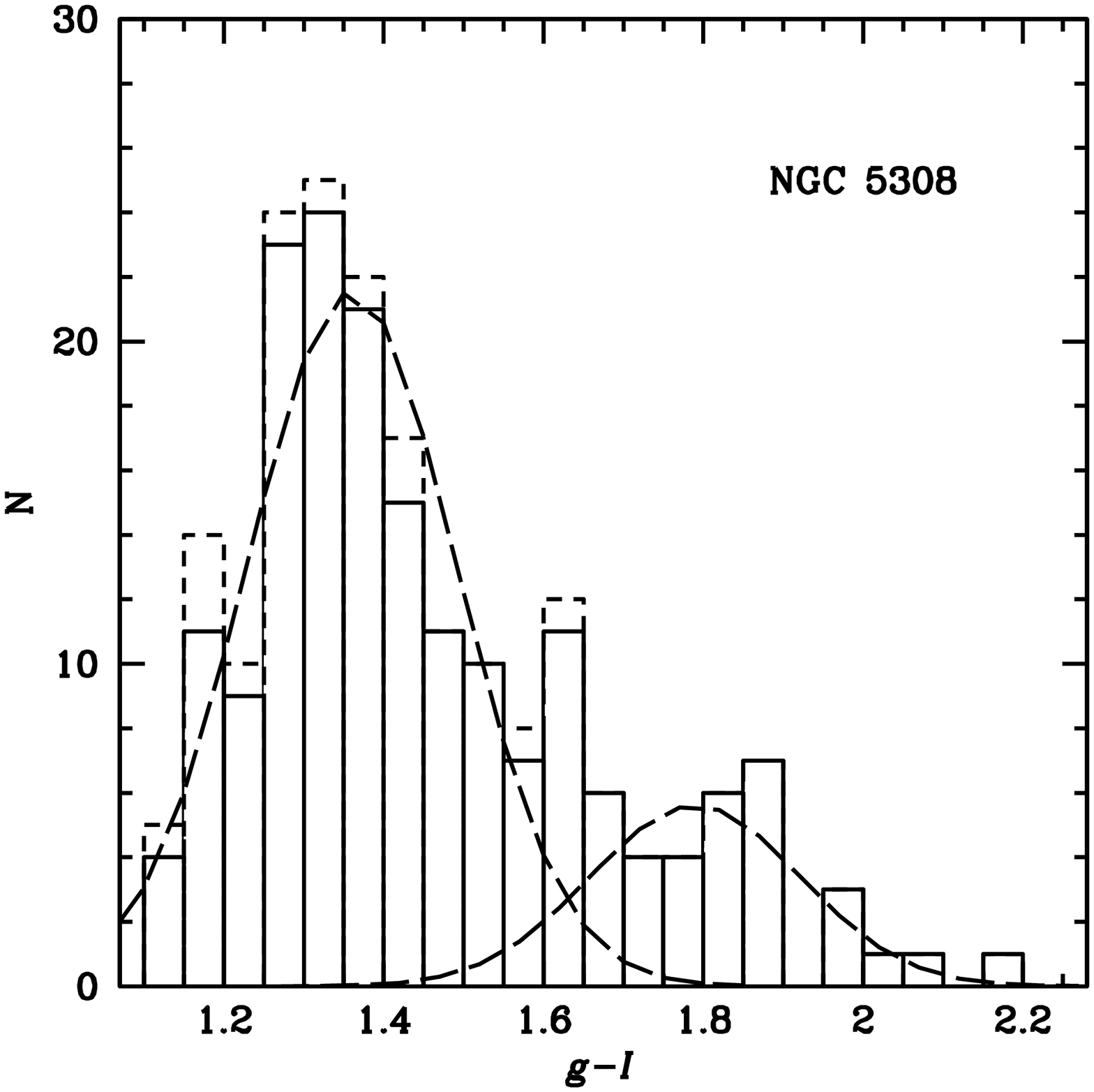}{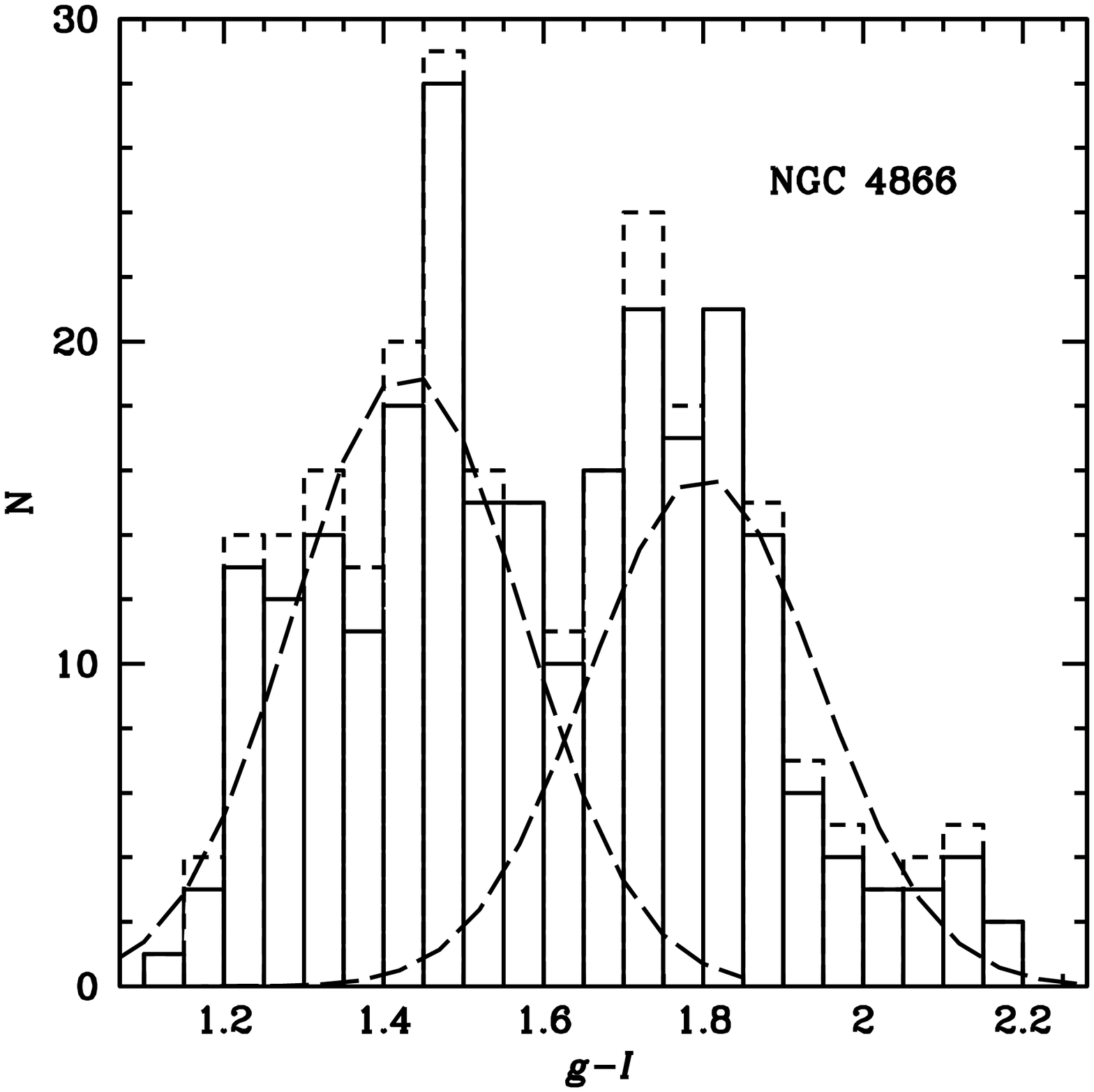}
\caption{GC color distributions for our target galaxy sample. The histograms with solid lines represent the color distribution
within the inner 14 kpc in each galaxy. The dashed curves are the estimates from
KMM while the dotted histograms represent the total number of clusters detected in the entire image.}
\label{colhist}
\end{figure*}

NGC~4710 is the nearest galaxy ($m-M$ = 31.0) while the other three are
more distant by a factor of 3--4. As a result, the ACS image of NGC~4710
covers the smallest radial extent ($\sim$14
kpc).
Therefore, we only consider GCs that lie within 14 kpc in all galaxies,
 and investigate whether or not
their color distributions are bimodal using the KMM algorithm
\citep{mclachlan88,ashman94}. 
We kept the dispersions the same for the two gaussians and found that 
the p-value ranged from
0.00 to 0.018. Low-p values reject the hypothesis that the examined
distribution resulted from a single Gaussian distribution.
Histograms of the $g-I$ color distributions within 14 kpc and the KMM fits are
shown in Fig.~\ref{colhist}. In all four target galaxies, we detect a dominating blue
GC population at \gI\ $\approx$ 1.4. This is consistent with the old, metal-poor
cluster population seen in normal elliptical galaxies.
We also clearly detect a
redder population at \gI\ $\approx$ 1.8 in the more luminous galaxies, NGC~5308 and
NGC~4866 and at \gI\ $\approx$ 1.6 in the less luminous galaxy, NGC~4710. NGC~5475
does not show a significant red GC population.
The peaks detected by the KMM algorithm are given in Table~1. The
red peak colors at \gI\ $\approx$ 1.8 and 1.6 are consistent
with the colors of old, metal-rich clusters
seen in elliptical galaxies with similar (bulge) luminosities \citep{larsen01,peng06}. 
Reddening does not seem to be significant in the galaxies and we do not expect that
any reddened ''blue" clusters were classified as ''red" since the KMM shows a clear
bimodality. Had the blue clusters been reddened considerably, we would have expected
to see more clusters in the valley between the two gaussians and not a clear bimodality.
A measurement of potential reddening in ellipticals
by \citet{gj95}, using IRAS (infrared) data and optical colors,
showed that any color gradient is small, with $B-I$ below 0.03 per dex.
These results for ellipticals suggest that reddening should not  
significantly affect the colors of clusters in early-type  
galaxies, such as the ones studied here.
\section{Specific Frequency of Globular Clusters}
\label{s:LFs}

We constructed the GC luminosity functions (GCLFs) for the galaxies within the
inner 14 kpc in the $I$-band after completeness corrections, and determined
turn-over magnitudes ($M_{I_{TO}}$) by means of a Gaussian fit. 

\begin{figure}
\plotone{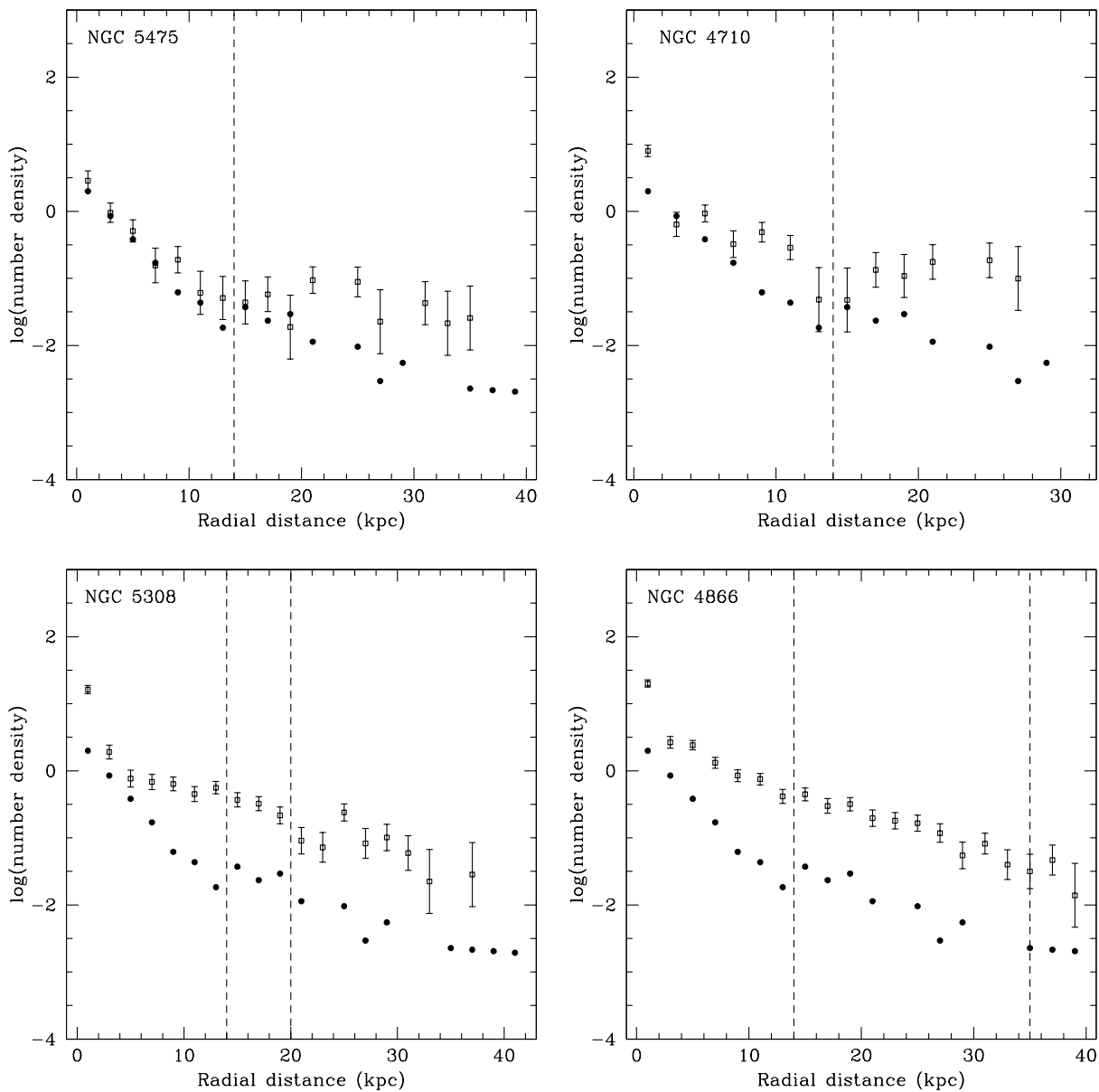}
\caption{Log of the surface number density per sq. kpc of the GCs as a function of radial distance from the center of the galaxy. Open squares and filled circles denote the values for the target galaxy and the Milky Way respectively. Vertical dashed lines are drawn at 14 kpc and at the extent of the GC system
of each target galaxy.}
\label{den}
\end{figure}

The total number of clusters within 14 kpc was computed by doubling the number
of observed 
clusters brighter than $M_{I_{TO}}$
\citep[cf.][]{harris81} for all galaxies except NGC~5308 for which we used $M_{I_{TO}}$=-8.46
since it showed a much brighter $M_{I_{TO}}$ (see Table~1.)
Correction for incomplete spatial coverage within 14 kpc was
done by assuming a symmetric cluster distribution.
Using $M/L$ ratios and the relation between the spatial extent of GCs
and the mass of the galaxy from \citet{rhode07}, we estimate the extent of their GC systems to be
$\sim$14 kpc (NGC~5475 and
NGC~4710), 20 kpc (NGC~5308), and 35 kpc (NGC~4866).  
It is clear from  Fig.~\ref{pos} that we do not sample the full azimuthal extent
of the GC system beyond a radius of 14 kpc in NGC~5308 and NGC~4866.
In Fig.~\ref{den} we show the number density (number of clusters per sq. kpc) vs the
radial distance from the center of the galaxy for each target galaxy for the
entire radial coverage in the ACS image. We have also added the number density
profile for the Milky Way for comparison. The bottom panels of Fig.~\ref{den}
show that the radial density profiles of GCs from 14 kpc to 20 kpc in
NGC~5308 and from 14 kpc to 35 kpc in NGC~4866
are similar to that in the Milky Way. Hence, we used a (scaled) extrapolation of the
GC system beyond 14 kpc from the Milky Way profile.

Using the radial distribution of the metal-poor and metal-rich Galactic GCs
as determined from the current version of the \citet{harris96} catalog, we
then estimate the total number of GCs in each of our sample galaxies and
calculate the total specific frequency ($S_N$) of the GC systems.
$S_N$ is defined as the number of star clusters per galaxy luminosity normalized
to an absolute {\it V\/} magnitude of $-$15 \citep{harris81}. 
We also compute $S_N$ for blue clusters alone.
See Table~\ref{t:results} for results.

\begin{figure*}
\plottwo{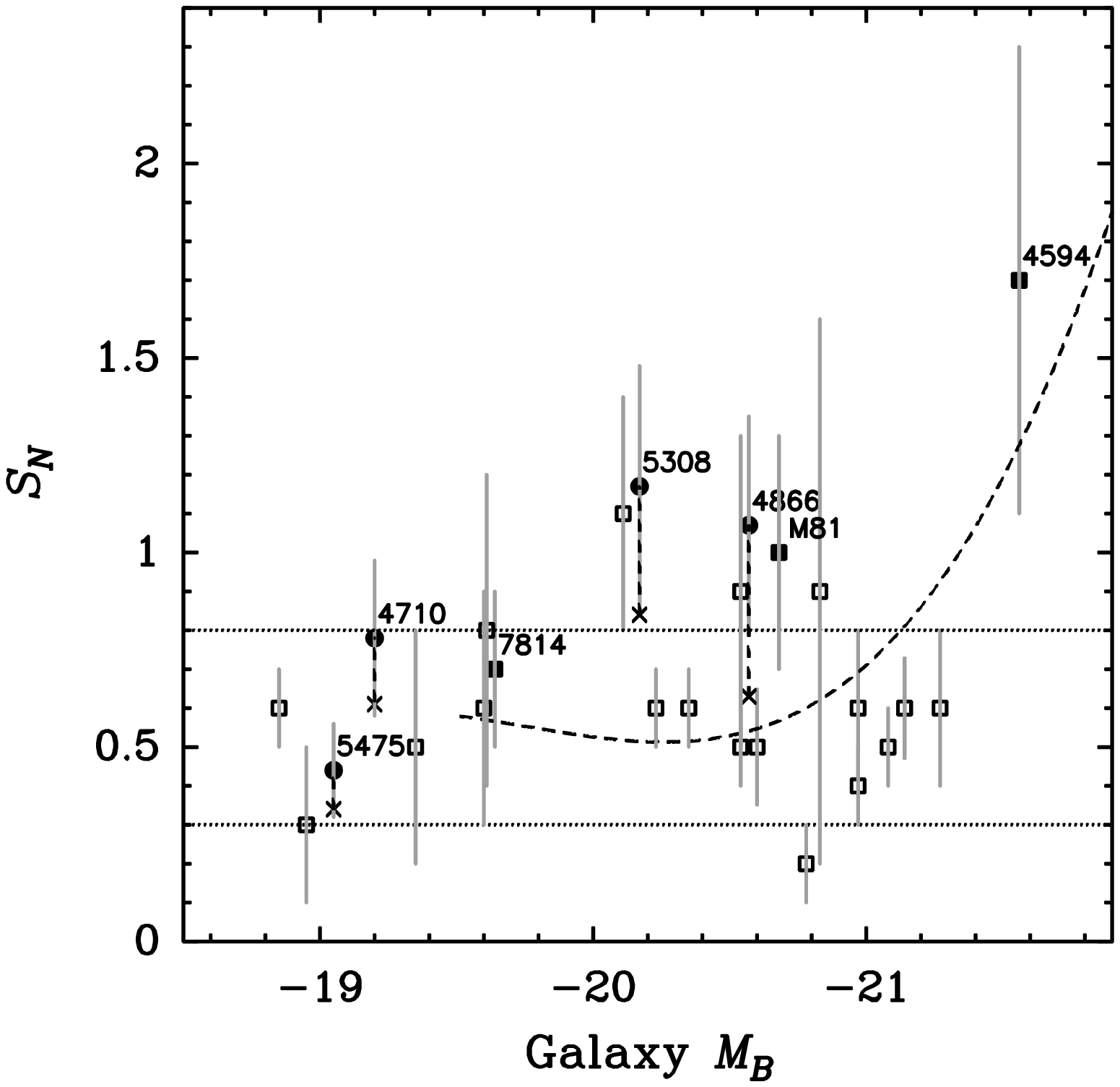}{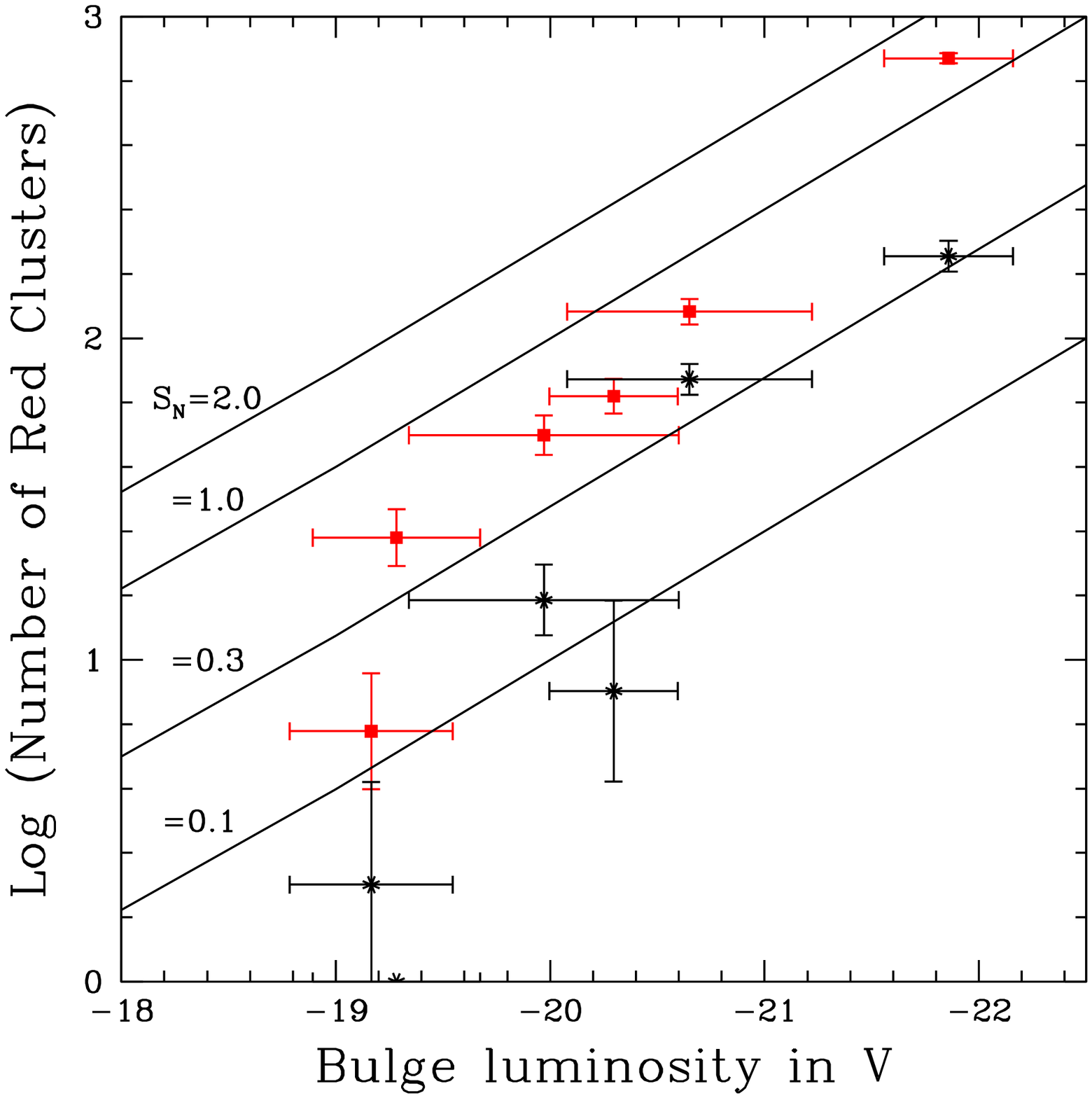}
\caption{{\it Left}: Relation between the galaxy $B$ luminosity and specific frequency. The filled circles denote the
total $S_N$ for the sample galaxies while the crosses are the $S_N$ values for the blue clusters alone.
The filled squares and the open squares show the total $S_N$ values for early-type and late-type
spirals respectively taken from \citet{goudfroo03,rhode07,chandar04}.
The dotted lines show the range of values seen for the total $S_N$ in late-type spirals \citep{goudfroo03} while the dashed line
depicts a least-square fit to the $S_N$ of blue clusters \citep{rhode07}. 
{\it Right}: The number of red clusters as a function of the bulge $V$ luminosity. The red filled squares
represent all the red clusters while the black asterisks represent the number of red clusters within 2 $R_{eff}$. The solid lines show
relations for $S_N$ =0.1, 0.3, 1.0, and 2.0.}
\label{snmb}
\end{figure*}

We could not find a $V$ magnitude for NGC~5475 in the literature so we
interpolated the SDSS {\it ugriz} magnitudes
and then converted the interpolated ABMAG to a Vega-based $V$ magnitude.
The left panel of Fig.~\ref{snmb} shows the relation between the total $B$ luminosity of each galaxy and $S_N$. For our sample galaxies, values for both
the total $S_N$ and the blue $S_N$ are shown. Total $S_N$ values for
early-type and late-type spirals from \citet{goudfroo03}, \citet{rhode07}, and
\citet{chandar04} are also shown.
We see that our blue $S_N$ values (0.34 -- 0.84) mostly lie within the typical
$S_N$ values for late-type spirals (0.55$\pm$0.25) found by these authors.
This strengthens the argument
that GC systems of spirals are made of a universal
halo-population of metal-poor GCs that are present in all galaxies.

\section{Discussion}

Here, we focus on using the properties of the red GCs to better
understand bulge assembly.
The right panel of Fig.~\ref{snmb} shows the relation between bulge luminosity
and the total number of red GCs (red symbols), as well as the number of red GCs located within
2$R_{eff}$ of the bulge (``bulge'' GCs; black symbols), for our sample
plus the Sa galaxies NGC~4594 \citep{spitler06}
and NGC~7814 \citep{goudfroo03}.
The total number of red GCs and bulge GCs clearly increase with bulge
luminosity. The superimposed lines of constant specific
frequency ($S_N$) show that the specific frequency does not remain
constant but changes with bulge luminosity.
Low-luminosity bulges (with $M_V$ $\ge -20$) have
lower values of $S_N$ than higher luminosity bulges.
Secular evolution leads to an increase in bulge luminosity
without a corresponding increase in the number of bulge clusters, since 
the low star formation rates associated with secular building of bulges 
\citep[e.g.,][]{sheth05} do not lead to the formation of clusters 
massive enough to survive a Hubble time \citep{bastian08}. Secular 
evolution therefore predicts no trend in the number of bulge clusters 
with bulge luminosity. On the other hand, bulge assembly through 
vigorous star formation events such as those occurring during mergers or 
interactions of gas-rich galaxies {\it is} known to be associated with 
the formation of massive star clusters \citep[e.g.,][]{whisch95,schw96,mill+97,
gou+01,gou+04,goudfrooij07}. 
Since the right panel of Figure 4 indicates that the number 
of bulge clusters increases with bulge luminosity (especially among the 
more luminous bulges), it seems fair to conclude that vigorous star 
formation events dominated secular evolution during the assembly history 
of those bulges.

Disruption of GCs, on both long and short timescales,
can potentially affect $S_N$.
We first consider the two disruption processes that have the most
significant impact on the long-term survival of  bulge GCs: 
Two-body relaxation (sometimes  
called ``evaporation'') and dynamical friction \citep[e.g.,][] 
{gneost97,fz01,vesp01,prieto08}. 
We only consider GCs brighter than or equal to the turnover
luminosity, corresponding to a mass of $\approx 2 \times 10^5$ M
$_{\odot}$ for an age of 12 Gyr \citep[e.g.,][]{mclfal08},
since these are the only clusters used in the $S_N$ calculation.
For evaporation, we use the simple parametrization of \citet{mclfal08}, which
evolves an initial Schechter function over time, at rates
that depend on the internal density of the clusters.
Differences in age have a small effect on $S_N$, at the 
$\approx10$\% level,
since for a typical density, the number of clusters
more massive than the turnoff decreases by only 12\%
between 1.5 and 12 Gyr, and by 9.8\% between
3 and 12 Gyr.
Based on the data presented in \citet{jordan05},
the sizes of red GCs 
are similar in the low and high-luminosity bulges of 
elliptical and S0 galaxies,
over a similar range of cluster luminosities.
This suggests that the internal densities, and hence the
evaporation rates, are also similar for (bulge) GCs 
in galaxies with very different bulge luminosities.
Over a Hubble time, dynamical friction will destroy GCs within 
two effective radii of the bulge if they are more massive
than ${\cal{M}}_{GC} \ga 3\times 10^{-6} {\cal{M}}_{\rm gal}$ (where ${\cal 
M}_{\rm gal}$ is the total mass of the host galaxy) \citep{gneost97}.
The largest effect will therefore occur for 
GCs more massive than ${\cal{M}}_{GC} \ga 1 \times 10^6$ M $_{\odot}$
in NGC~5475, our lowest-luminosity
galaxy, with $M_V = -19.72$ (${\cal{M}_{\rm gal}}$ $\simeq$  $3 \times 10^{11}$ M$_{\odot}$ assuming
${\cal{M}}/L_V = 6.1$ for an Sa galaxy \citep[e.g.,][]{fabgal79}).
Clusters this massive only represent a fraction $\la 0.09$ of a typical
GC mass function.
Therefore, we do not expect processes that affect GCs on long
timescales to impact the 
$S_N$ values plotted in Fig.\ 4 at much more than the 10\% level,
which is smaller than the typical uncertainty in $S_N$.

Processes that disrupt clusters on shorter ($\tau \lesssim 10^9$~yr)
timescales are not as well understood as
evaporation and dynamical friction, nor is 
it known whether the processes which disrupt young clusters
in galaxies today are similar to those experienced by
GCs during their youth.
There is however, recent evidence which suggests that the early
disruption of clusters in nearby galaxies
does not vary strongly from galaxy to galaxy \citep[e.g.,][]{fall09,chandar10},
because it is dominated mainly
by processes internal to the clusters themselves.
If this was also true for the early evolution of GCs,
then disruption processes in general should not have
impacted $S_N$ differently from galaxy to galaxy.

\begin{figure}
\plottwo{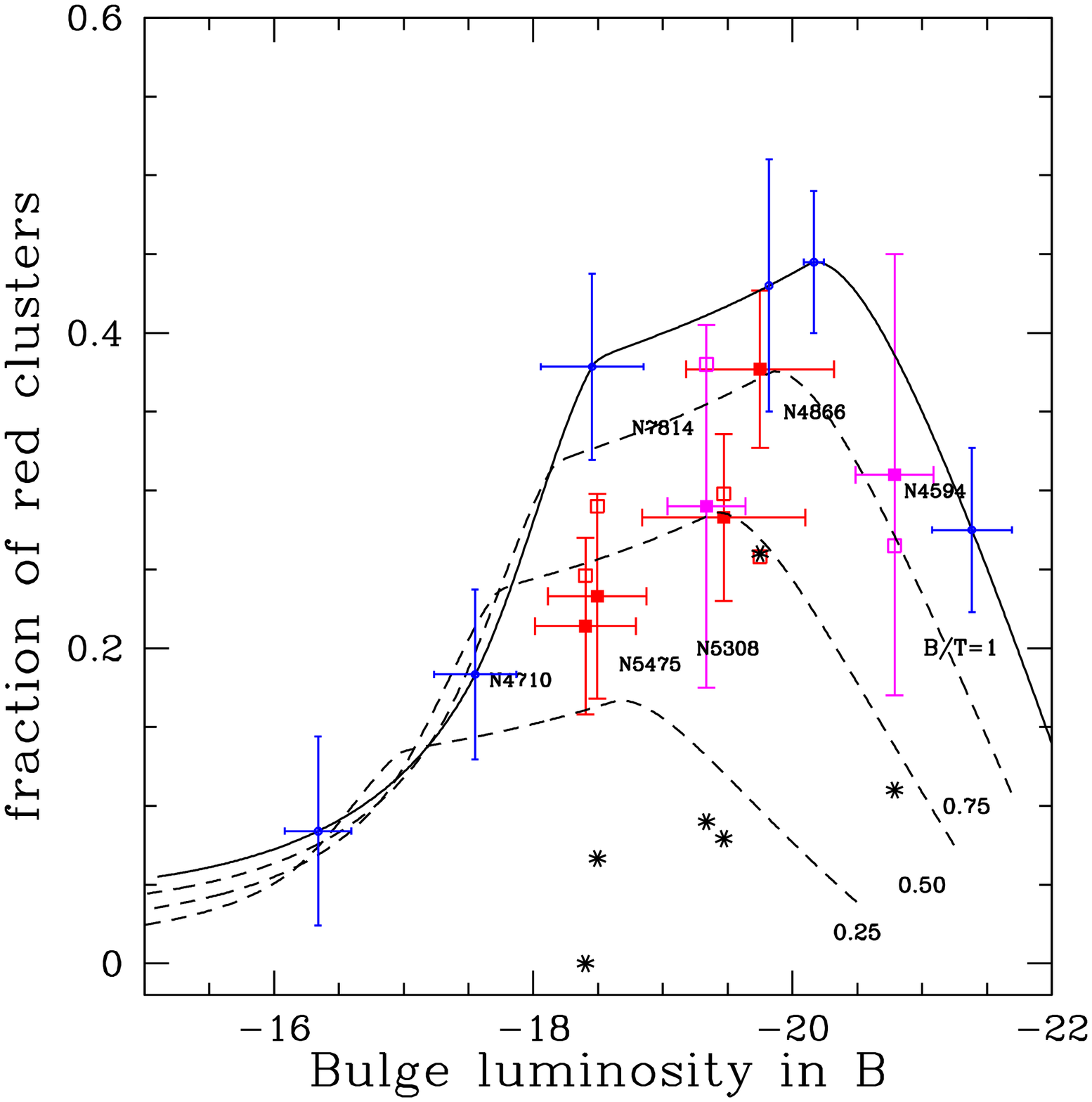}{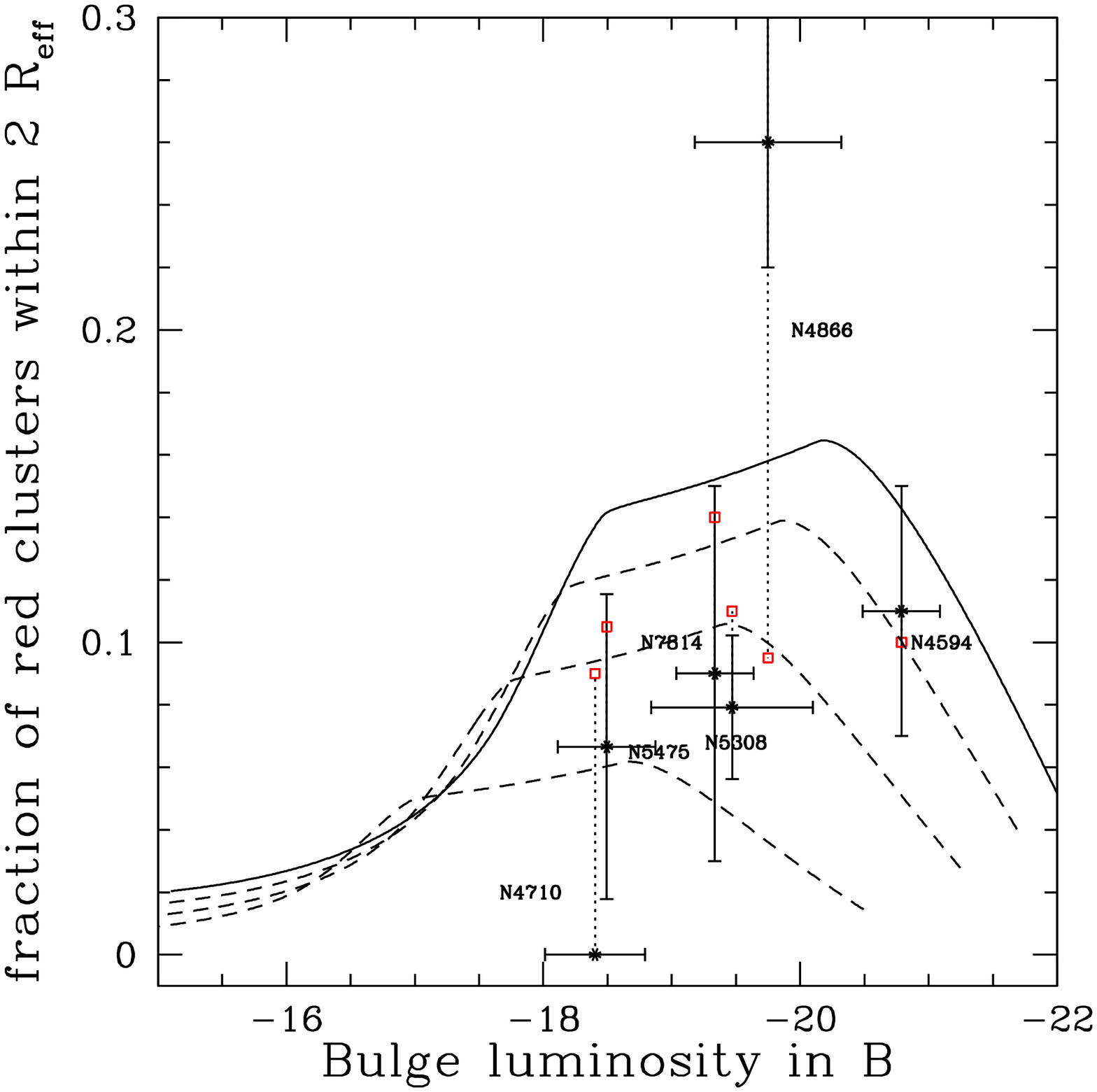}
\caption{{\it Left}: Relation between the bulge luminosity in $B$ and the fraction of red GCs. The solid black curve is a
fit to the ellipticals (blue circles) from \citet{peng08}. Black dashed lines represent the expected relation between the bulge luminosity
and the fraction of red clusters for $B/T$ = 0.75, 0.50, and 0.25 if all the red clusters formed in a
manner similar to that in ellipticals. Red and magenta filled squares are the observed values for the
target galaxies and comparisons from \citet{goudfroo03} and \citet{spitler06}, respectively.
Open red squares indicate the
predicted fraction of red GCs based on their $B/T$ values. Black asterisks show the fraction of red GCs within 2 $R_{eff}$. {\it Right}: Relation between the bulge luminosity in $B$ and the fraction of red GCs within 2 $R_{eff}$. Black asterisks show the fraction of red GCs within 2 $R_{eff}$ while the red open squares show their predicted fractions based on their
$B/T$ values.}
\label{fr}
\end{figure}

The left panel of Fig.~\ref{fr}
shows the fraction of red clusters ($f_{\rm red} = N_{\rm red}/(N_{\rm
red}\,+\,N_{\rm blue}))$ versus bulge luminosity.
The solid line shows an empirical relation between
$f_{\rm red}$ and bulge luminosity for elliptical galaxies, based on 
the compilation of \citet{peng08}, and provides a prediction of
$f_{\rm red}$ when bulges form purely via dissipative  
merging. $f_{\rm red}$ increases with bulge luminosity
and then drops off, 
likely due to the fact that the
$S_N$ of blue GCs increases with increasing galaxy luminosity in $B$. \citep{rhode05}.
Assuming that the number of blue GCs in spirals  
scales with the total luminosity of the host galaxy, there should be a  
factor of $1/(B/T)$ more blue GCs per unit bulge luminosity in spirals
relative to ellipticals.
Hence, the expected fraction of red GCs in spiral galaxies with  
different $B/T$ ratios in this dissipative scenario
is given by $f_{\rm red, diss} = N_{\rm red}/ (N_{\rm red} + N_{\rm
blue} \times 1/(B/T))$. $f_{\rm red, diss}$ is shown for three
different $B/T$ ratios (0.75, 0.50, and 0.25) with dashed lines in
Figure~\ref{fr}. Open squares mark values of $f_{\rm red, diss}$ 
predicted red GC fractions based on the observed $B/T$ ratios of
our sample spirals.  The right panel of Figure~\ref{fr} is a similar
plot but for red clusters within 2 $R_{eff}$ of the galaxy bulges,
which we derive to be a factor of 0.37$\pm$0.10 lower than the total
red fraction, using
$R_{eff}$ values from \citet{faber89} and
GC data from \citet{jordan09} and \citet{peng08}.

From Fig.~\ref{fr} we observe the following trends: The fraction of
{\em all} red GCs shows a trend similar to that seen in ellipticals (solid
curve).  However, comparing the fraction of {\em all} red GCs to that of
{\em bulge} GCs
reveals some intriguing results. Spirals
with faint bulges ($M_B>-19.5$ mag) tend to have fractions
of red bulge GCs (asterisks in Fig.~\ref{fr}) that fall
significantly below the values predicted from their $B/T$ ratios (open
squares), i.e. below the
values expected from purely dissipative processes.
However, the fraction of bulge GCs are generally
consistent with those expected from merging processes for galaxies
with higher luminosity bulges, NGC~4866 and NGC~4594. 
NGC~4866 has a relatively
large $R_{eff}$, and the high value observed for the bulge
clusters in the right panel of Fig~\ref{fr} is likely due to the
inclusion of some disk clusters.

\section{Conclusions}
We have studied four edge-on Sa galaxies and investigated the properties of
their GC systems and their relation to their host galaxy
bulges. We find that the total number of red clusters as well as the
red clusters associated with the bulge in Sa spirals increase with
increasing bulge luminosity.
However, the bulge $S_N$ of lower luminosity
Sa's are a factor of 2--3 lower than those of higher luminosity spirals.
 We also find that all galaxies in our sample host a
blue, presumably old metal-poor GC system with estimated specific
frequencies ranging from 0.34 to 0.84, consistent with
typical values seen in late-type spirals. We detect a number of red disk
clusters in all the sample galaxies. Our results suggest a
picture where dissipative processes were more important for
building luminous bulges of Sa galaxies, whereas secular
evolution may have played a larger role in building lower-luminosity
bulges in spirals. More $HST$ observations of GC systems in spiral galaxies
are needed to verify these results.

\acknowledgements
We thank the anonymous referee for carefully reading our manuscript, and for
suggestions which helped to improve this
paper. Support for this work was provided by NASA through HST grant number GO-10594 from
the Space Telescope Science Institute, which is operated by the Association of
Universities for Research in Astronomy, Inc., under NASA contract NAS5-26555.
THP acknowledges support in form of the Plaskett Research Fellowship  
at the Herzberg Institute of Astrophysics of the National Research  
Council of Canada

{}

\newpage

\begin{deluxetable}{lcccc}
\centering
\tabletypesize{\scriptsize}
\tablecaption{Basic properties of the galaxy sample\label{t:results}}
\tablewidth{0pt}
\tablehead{\colhead{} & \colhead{NGC~5475} &
\colhead{NGC~4710} & \colhead{NGC~5308} & \colhead{NGC~4866}}
\startdata
m-M&32.3&31.0&32.4&32.3\\
$M_{B_T^0}$&-19.05&-19.20&-20.17&-20.57\\
$M_{K}$&-22.89&-23.43&-24.04&-24.38\\
$(g-I)_{KMM}$ &1.21,1.53&1.37,1.64&1.36,1.79&1.43,1.80\\
$m_{I_{TO}}$&23.99$\pm$0.09&22.65$\pm$0.17&23.41$\pm$0.06&24.13$\pm$0.14\\
$M_{I_{TO}}$&-8.30&-8.35&-8.99&-8.17\\
$n_{bulge}$&2.9&1.2&3.5&3.0\\
$S_N$&0.44$\pm$0.12&0.78$\pm$0.20&1.17$\pm$0.31&1.07$\pm$0.28\\
$S_N$(blue)&0.34$\pm$0.09&0.61$\pm$0.16&0.84$\pm$0.22&0.63$\pm$0.16\\
red fractn.&0.23$\pm$0.06&0.21$\pm$0.06&0.28$\pm$0.05&0.38$\pm$0.05\\
B/T&0.60&0.48&0.53&0.47\\
$(g-I)_{\rm div}$ & 1.50 & 1.50 & 1.55 & 1.60 \\
\enddata
\tablecomments{Row~(1): Object ID. Row~(2): Distance moduli calculated using
velocity from http://nedwww.ipac.caltech.edu (NED) and assuming ${\it
    H}_{0}$ = 70 km s $^{-1}Mpc^{-1}$. Row~(3): Absolute $B$
  magnitude from NED. Row~(4): Absolute $K$ magnitude from NED. Row~(5):
$g-I$ values of the peaks detected by KMM within 14 kpc.
Row~(6): The magnitude at the turn-over of the
luminosity function in the $I$ band. Row~(7): The absolute magnitude of the
turn-over in the $I$-band. Row~(8): S\'{e}rsic index of the bulge. Row~(9): The total
specific frequency for the GCs. Row~(10): The specific frequency of the blue clusters alone.
Row~(11): The fraction of the red clusters obtained by dividing the number of red clusters by the total
number of clusters. Row~(12): Bulge to total luminosity ratio of the
galaxy in the $I$-band. Row~(13): \gI\ color to divide ``blue'' from ``red'' clusters.}
\end{deluxetable}

\end{document}